\documentclass[12pt,a4paper]{article} 
\usepackage{psfig,latexsym}

\newlength{\figwidth}
\newcommand{\lsed}{\mbox{$\ell_{\mbox{\scriptsize SED}}$}}
\newcommand{\kb}{\mbox{$\mbox{k}_{\mbox{\scriptsize B}}$}}
\newcommand{\ev}{\mbox{eV}}

\begin{document}
\begin{center}
{\large \bf
Unconventional MBE Strategies from Computer Simulations for
 Optimized Growth Conditions}

{\sc S. Schinzer\raisebox{0.8ex}{\rm \small a}\footnote{Corresponding
  author. FAX:+49-931-888-4604; e-mail:
  schinzer@physik.uni-wuerzburg.de}, 
M. Sokolowski\raisebox{0.8ex}{\rm \small b}, 
M. Biehl\raisebox{0.8ex}{\rm \small a}, and
  W. Kinzel\raisebox{0.8ex}{\rm \small a}}

{\em 
\raisebox{0.8ex}{\rm a}Theoretische Physik, TP III,
\raisebox{0.8ex}{\rm b}Experimentelle Physik, EP II\\
  Universit\"at W\"urzburg, Am Hubland, D-97074 W\"urzburg, Germany
}
\end{center}

\begin{abstract}
We investigate the influence of step edge diffusion (SED) and
desorption on Molecular Beam Epitaxy (MBE) using kinetic Monte--Carlo
simulations of the solid--on--solid model.  Based on these
investigations we propose two strategies to optimize MBE growth.  The
strategies are applicable in different growth regimes: During
layer--by--layer growth one can exploit the presence of desorption in
order to achieve smooth surfaces. By additional short high flux pulses
of particles one can increase the growth rate and assist
layer--by--layer growth. If, however, mounds are formed
(non--layer--by--layer growth) the SED can be used to control size and
shape of the three-dimensional structures. By controlled reduction of
the flux with time we achieve a fast coarsening together with smooth
step edges.
\end{abstract}


\section{Introduction}
The growth of high quality compound semiconductors is of great
technological importance \cite{hs96}. Despite the longstanding
tradition of molecular beam epitaxy (MBE), it is still a challenging
task to improve the growth of high quality thin films and well defined
interfaces. In order to optimize MBE growth a detailed knowledge of
the relation between microscopic processes and macroscopic properties
is very important. Computer simulations are an ideal tool to access
this relation between atomistic processes and epitaxial growth. In
addition, different and new growth strategies can be easily
implemented and tested in a fast and cheap way \cite{rlw95,js98}.

In this paper we will investigate the macroscopic effects of two
distinct microscopic mechanisms. The term {\em microscopic} refers to
processes on the atomic scale: e.g. a single diffusion step of an
adatom or desorption of an atom. These processes are the ingredients
to the computer model used in this paper. This is contrasted to the
term {\em macroscopic} for effects which are typically measurable in
experiments: e.g. the overall mass desorption as can be monitored by
the partial pressure \cite{jfh90}, the form and the distribution of
three--dimensional structures accessible by scanning tunneling
microscopy \cite{oes96}, or the growth rate as determined by electron
diffraction oscillations \cite{umm91}.  The computer simulations
employed here are ideally suited to bridge the gap between such
macroscopic effects and their underlying microscopic processes, since
both scales are accessible.

Several strategies have been proposed in the literature to optimize
MBE growth: In particular, layer--by--layer growth is most desirable
in order to achieve high quality thin films
\cite{rlw95,js98}. However, quite often a transition to
non--layer--by--layer is observed where three--dimensional (3D)
structures such as mounds or pyramids appear. In {\em conventional}
MBE \cite{note1} the time $t_\times$ until this growth mode crosses
over to 3D--growth has been shown to vary with $ F t_\times \approx
(D/F)^\delta$ \cite{kbk97}, where $F$ stands for the flux and $D$ for
the diffusion constant of adatoms. Without desorption,
Ehrlich--Schwoebel barriers and step edge diffusion (SED) $\delta =
2/3$ has been observed for epitaxial growth \cite{kbs99}. For metals,
several methods have been proposed and tested to achieve and maintain
layer--by--layer growth. For instance it has been shown that pulsing
the deposition rate or pulsing the temperature leads to a prolonged
layer--by--layer regime \cite{rlw95}. Recently, it has been proposed
that pulsed glancing--angle sputtering can even lead to
``layer--by--layer growth forever'' \cite{js98}. All these concepts
can so far be understood in terms of a typical diffusion length or an
enhanced interlayer diffusion at step edges.

In this paper we will propose strategies which exploit other specific
microscopic processes, namely desorption \cite{jlp97} and SED
\cite{af96c,ssw97a}.  As far as we know, no attempt 
has been made to exploit these processes in order to achieve improved
growth. Some preliminary results of our investigation have been
published in \cite{ssb98}, and in this paper we describe the
investigation in full detail.

In Sec.\ \ref{model} we introduce the solid--on--solid (SOS) model and
the microscopic processes. In our computer experiments we first
investigated the temperature dependence of the overall growth rate in
the layer--by--layer regime (Sec.\ \ref{des}).  We are able to
correlate this macroscopic property to the microscopic dynamics of the
computer model. This allows us to propose a new strategy for
layer--by--layer growth.  If, however, the growth of
three--dimensional structures occurs another strategy is
applicable. Using a simplified model of growth we have recently shown
that SED plays a crucial role in this regime \cite{bks98,skb98}. These
findings allow us to propose an optimized way for the growth of
3D--structures in Sec.\ \ref{sed}. Concluding remarks concerning the
experimental realisation and a summary will be given in Sec.\
\ref{conclusion}.

\section{Computational model}              \label{model}
Lattice models with the solid--on--solid restriction (SOS) have been
proved to be a useful tool to study surface morphology
\cite{bar95,vp95}. The model has a long history for the study of the
surface roughness transition \cite{wee80}. Gilmer and Bennema were the
first (to our knowledge) who included surface diffusion
\cite{gb72}. Since then it has been intensively used to study
epitaxial growth \cite{lk97,kru97b}.

Here we use its most simple form where only one kind of particles and
a simple cubic lattice is considered. The particles represent single
atoms when a comparison with a simple cubic metal is made. However,
even compound semiconductors can be modeled, as long as kinetic
features are investigated only. E.g. in Ref.\ \cite{sv93} the RHEED
(reflection high energy electron diffraction) oscillations of
GaAs(001) during growth have been quantitatively reproduced.

In our simulations we use the Maksym--algorithm of Ref.\
\cite{mak88}. At each time step a Monte--Carlo move is carried
out. The way how the event is selected makes it superior to
conventional Monte--Carlo techniques (the algorithm uses partly a
binary search in the array of possible events).  We have used a system
of 300 times 300 lattice sites, if not stated otherwise.

Besides the SOS--restriction further simplifications are due to the
particular choice of possible events labeled $i$ and the
parametrisation of the corresponding rates $\Gamma_i$. We allow jumps
to the four nearest neighbor sites (diffusion on a flat surface,
attachment and detachment from steps, \ldots) and desorption. The
rates do only depend on the four nearest neighbor sites as will be
described below. We describe all these processes as
Arrhenius--activated
\begin{equation}
\Gamma_i = \nu_i \exp \left(-\frac{E_i}{\kb T} \right),
\end{equation}
as is predicted by several theories \cite{pon90}.

One quite often assumes vibration frequencies $\nu_i$ of the order of
Debye--frequencies, i.e. $10^{12}$ -- $10^{14} \mbox{s}^{-1}$. Indeed,
in sublimation experiments of CdTe(001) $10^{14} \mbox{s}^{-1}$ has
been observed \cite{jfh90,blw95}; vibration frequencies for diffusion
are often of the order of $10^{12} \mbox{s}^{-1}$ (measurements for
metals \cite{kel94}, calculations for GaAs(001) \cite{ks96ip,ohn96},
or simulations and calculations for Si(001) \cite{ald96,mkw91}).
Hence, it is reasonable to assume that the diffusion as well as the
desorption rates of our model share one common prefactor $\nu_i =
\nu_0 = 10^{12} \mbox{s}^{-1}$ which allows to keep the number of
parameters small.

The activation energy for the different microscopic processes are
parameterized as follows: a diffusion jump of a free adatom has to
overcome a barrier $E_B$, each next in--plane neighbour adds an energy
$E_N$. The rate of diffusion jumps which keep the height of the
particle unchanged thus becomes $\nu_0 \exp \left(- (E_B+n E_N)/ \kb T
\right)$, where $n$ represents the number of next in--plane
neighbors. Note, that the overall rate for diffusion on a flat surface
is four times this jump rate due to the four possible
directions. Hence, the diffusion constant becomes $D = \nu_0 \exp
(-E_B /\kb T)$ \cite{zan88}.  Since we measure all length scales in
units of the lattice constant $a$ we have neglected the term
$a^2$ in $D$. At step edges an additional Ehrlich--Schwoebel barrier
$E_S$ is considered \cite{eh66,ss66}. However, this barrier is not
added for particles on top of an elongated islands of only one lattice
constant width \cite{pkn91}. The desorption barrier is $E_D$. Again,
each next in--plane neighbour contributes $E_N$.

The deposition of particles occurs with a rate $F$ measured in
monolayers per second (ML/s). During deposition we consider another
process which is not Arrhenius--activated. After a deposition site is
chosen randomly we allow the particle to relax to a lower neighboring
site. Here, we consider only relaxation to nearest--neighbor sites. Such
{\em transient diffusion} or {\em downward funneling} has been 
observed in molecular dynamics of simple Lennard--Jones systems
\cite{est90,yhp98} and has been related to the reentrant
layer--by--layer growth at very low temperatures \cite{swv93-1}. In
addition it has been shown to play a crucial role for slope selection
in mound morphology \cite{bks98}.

We will concentrate on one set of parameters, namely $E_B=0.9 \ev$,
$E_N=0.25 \ev$, $E_S = 0.1 \ev$, and $E_D = 1.1 \ev$. This particular
choice of parameters reproduces some features of CdTe(001) during
sublimation \cite{sk98,nsk99} and annealing. However, we would like to
stress that the findings of our present work are of more general
relevance, independent of the specific choice of the energetic
parameters.

\section{Reevaporation during layer--by--layer growth
  and the Flush Technique}      \label{des}
For clarity we will distinguish several processes of
desorption. The term {\em desorption} will be explicitly used to
describe the atomistic process: the desorption of a single atom. {\em
  Sublimation} is reserved to describe the evaporation of a surface
when left in (perfect) vacuum. {\em Reevaporation} or more precisely
{\em reevaporation during growth} will describe the overall
desorption rate mostly due to the desorption of freshly deposited
particles during growth.

\begin{figure}[t]                  
\psfig{file=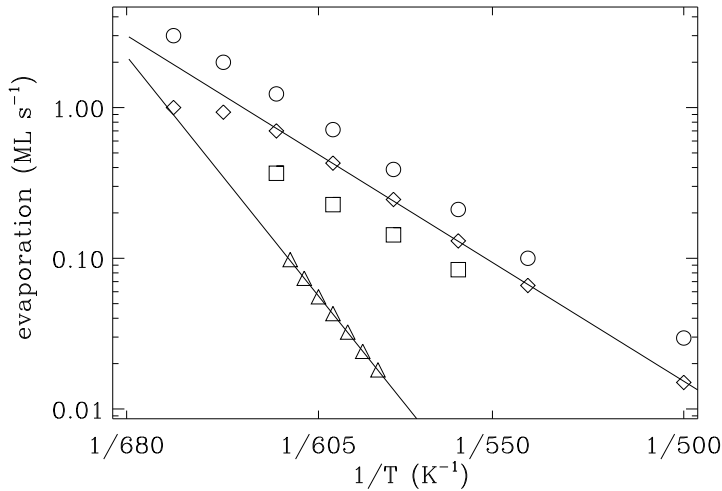,width=\figwidth}
\caption{ Reevaporation rate during growth with $F=$ 1
  ML/s ($\Diamond$), $F=$ 4 ML/s ({\footnotesize $\bigcirc$}), and
  sublimation rate $\times 10$ at $F=0$ ML/s ($\bigtriangleup$). In
  addition we show the reevaporation rate using the proposed
  flush-technique ($\Box$) with a mean flux of 1 ML/s consisting of a
  constant flux of 0.77 ML/s plus an additional pulse of 0.23 ML
  during 0.003 s at the beginning of each
  second. \label{des_arrhenius} }
\end{figure}

For two of the most important compound semiconductors a decrease of
the MBE growth rate with increasing temperature was observed
(CdTe(001) \cite{umm91,ac95,blw95} and GaAs(001)
\cite{tna97,tnz97}). For CdTe(001) the reevaporation rate was found to
follow an Arrhenius--rate with considerably lower values of the
activation energy of 0.14 -- 0.30 \ev\ compared to sublimation (1.55
\ev\ resp. 1.9 \ev\ \cite{nsk99}). A tempting explanation is to
ascribe this low energy to the existence of a physisorbed precursor
\cite{blw95}. However, studies of the sublimation with computer
simulation \cite{sk98} as well as experiments for CdTe(001)
\cite{nsk99} showed a strong influence of the morphology. In Ref.\
\cite{sk98} we already concluded that in MBE, one should expect
desorption rates other than those measured by
sublimation. Independently Pimpinelli and Peyla also showed that a
physisorbed precursor is not necessary to explain the observed low
energies using kinetic Monte--Carlo simulations as well as simple
scaling arguments \cite{pp98,ppc98}.

In fig.\ \ref{des_arrhenius}, the diamonds  represent the reevaporation
which we derived from the difference
between the applied flux ($F = 1$ ML/s) and the measured growth rate
({\em i.e}.\ the reached height/simulated time). The data points for $F
= 4$ ML/s ({\footnotesize $\bigcirc$}) show that the effective energy
is independent of the applied flux. The triangles
correspond to the sublimation {\em i.e}.\ the evaporation rate
without application of an external flux ($F=0$ ML/s) \cite{sk98}. Both
processes are found to be Arrhenius--activated, however, with strikingly
different effective energies. The reevaporation rate during growth
corresponds to an activation energy of approximately 0.90 \ev\ which
is even lower 
than the microscopic desorption energy of $E_D =$ 1.1 \ev. At high
temperatures the reevaporation rate saturates and equals the flux of
impinging particles. On the contrary, the sublimation energy of
approximately 1.73 \ev\ is considerably higher \cite{note2}.

\begin{figure}[t]
\psfig{file=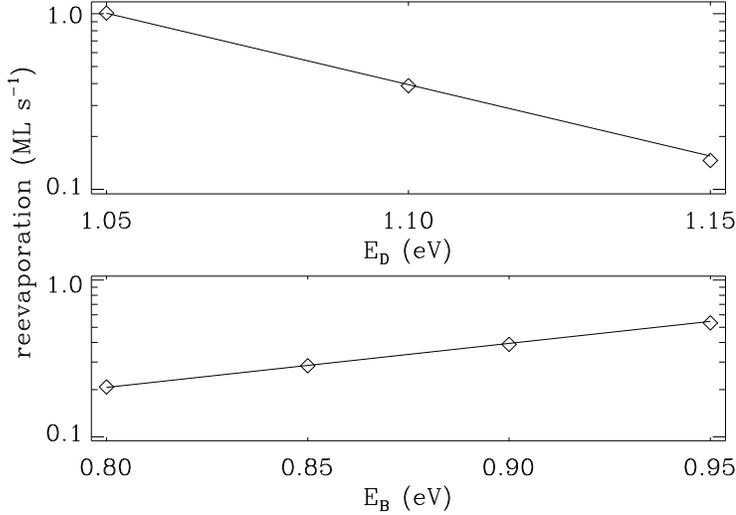,width=\figwidth}
\caption{ \label{mikros_ana} 
  Variation of the reevaporation rate during growth as a function of
  the desorption barrier $E_D$ (upper) and diffusion barrier $E_B$
  (lower curve). The slope gives the contribution to the effective 
  energy. The simulations were carried out at $T=560$ K with $F=4$
  ML/s.  }
\end{figure}

The relation of the sublimation energy to the microscopic parameters
was shown to be approximated by $E_{\mbox{\scriptsize sub}} \approx
0.61 E_B + 0.35 E_D + 2.85 E_N + 0.44 E_S$ \cite{sk98}. To derive this
relation we varied all microscopic energy parameters independently.
Applying the same microscopic analysis to the reevaporation during growth
of this model we obtain
\begin{equation}  \label{edes}
  E_{\mbox{\scriptsize re}} \approx -0.31 E_B + 0.94 E_D + 0.51 E_N -
  0.03 E_S.
\end{equation}

As an example for this microscopic analysis, fig.\ \ref{mikros_ana} shows
the measured influence of the diffusion barrier $E_B$ and of the
desorption barrier $E_D$ to the reevaporation rate. We applied a flux
of $F = 4$ ML/s. Since the measured reevaporation rate is much lower
(less than 1 ML/s) we can be sure to have no saturation effects. Note
the opposite sign of the two contributions. The slope measures the
prefactor in the above expression of $E_{\mbox{\scriptsize re}}$
({\it eq}.\  \ref{edes}).

We want to mention that this result does not agree with the scaling
relation obtained by Pimpinelli and Peyla \cite{pp98}. However, at
lower temperatures (not shown) we observe a cross--over to their
result with a critical nucleus size of $i^*=1$. The crossover itself
can be seen at the data point for $F = 4$ ML/s at 500 K which lies
above the value of about 0.2~ML/s which is extrapolated at 500 K from
data points at higher temperatures. A detailed 
investigation of the validity--regime of our result and the relation
to the results obtained by Pimpinelli and Peyla will be postponed to
a future work.

Besides the different weightings in $E_{\mbox{\scriptsize re}}$ and
$E_{\mbox{\scriptsize sub}}$ the striking difference (at high as well
as low temperatures) is the negative contribution of the diffusion
barrier $E_B$ to $E_{\mbox{\scriptsize re}}$. This result seems to be
of general validity \cite{pp98} and can be explained in the following
way: Even though the island distance is influenced by $E_B$, the
dominant effect of higher diffusion barriers seems to be the reduction
of the diffusion length of the adatoms. Consequently, particles have a
higher probability for desorption before they stick to an island.

\begin{figure}[t]
\psfig{file=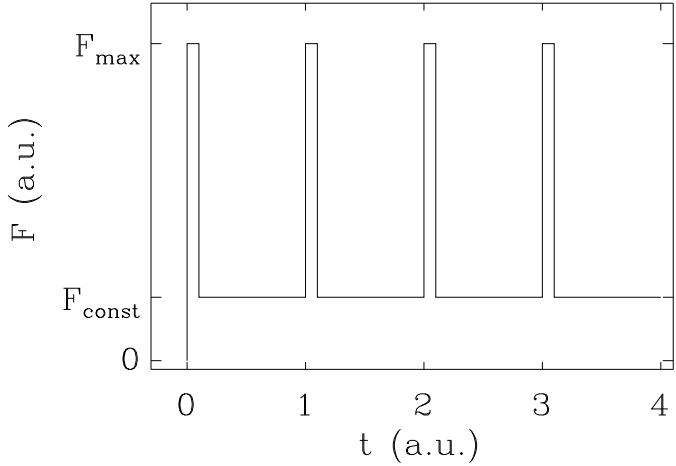,width=\figwidth}
\caption{ \label{skizze1}
  Schematic variation of the flux using the flush--technique.
}
\end{figure}

This result suggests a strategy to obtain high-quality
(layer-by-layer) growth together with high growth rates. Short flushes
of particles at the beginning of each monolayer would result in a
great density of islands. Afterwards with a
low flux the particles probably hit islands to stick to which will
result in a low overall reevaporation rate. The proposed procedure
(flush--mode) is drawn schematically in fig.\ \ref{skizze1}.

Figure \ref{des_arrhenius} shows that the reevaporation rate indeed
reduces by a factor of about two when applying this strategy. The mean
flux was 1 ML/s as for the conventional growth simulations. The
profile of the flux was composed as follows: In intervalls of one
second we deposited a total amount of 0.23 ML within 0.003 s
(see fig.\ \ref{skizze1}). Afterwards a constant flux of 0.77 ML/s was
applied. According to the decrease of evaporation the growth rate
increases. The gain is highest at high temperatures (at 620 K the
growth rate is doubled) since there the evaporation rate becomes
comparable to the applied flux.

\begin{figure}[t]
\setlength{\unitlength}{\figwidth}
\begin{picture}(1,0.707)(0,0)
\put(0,0){\parbox[b]{\figwidth}{\psfig{file=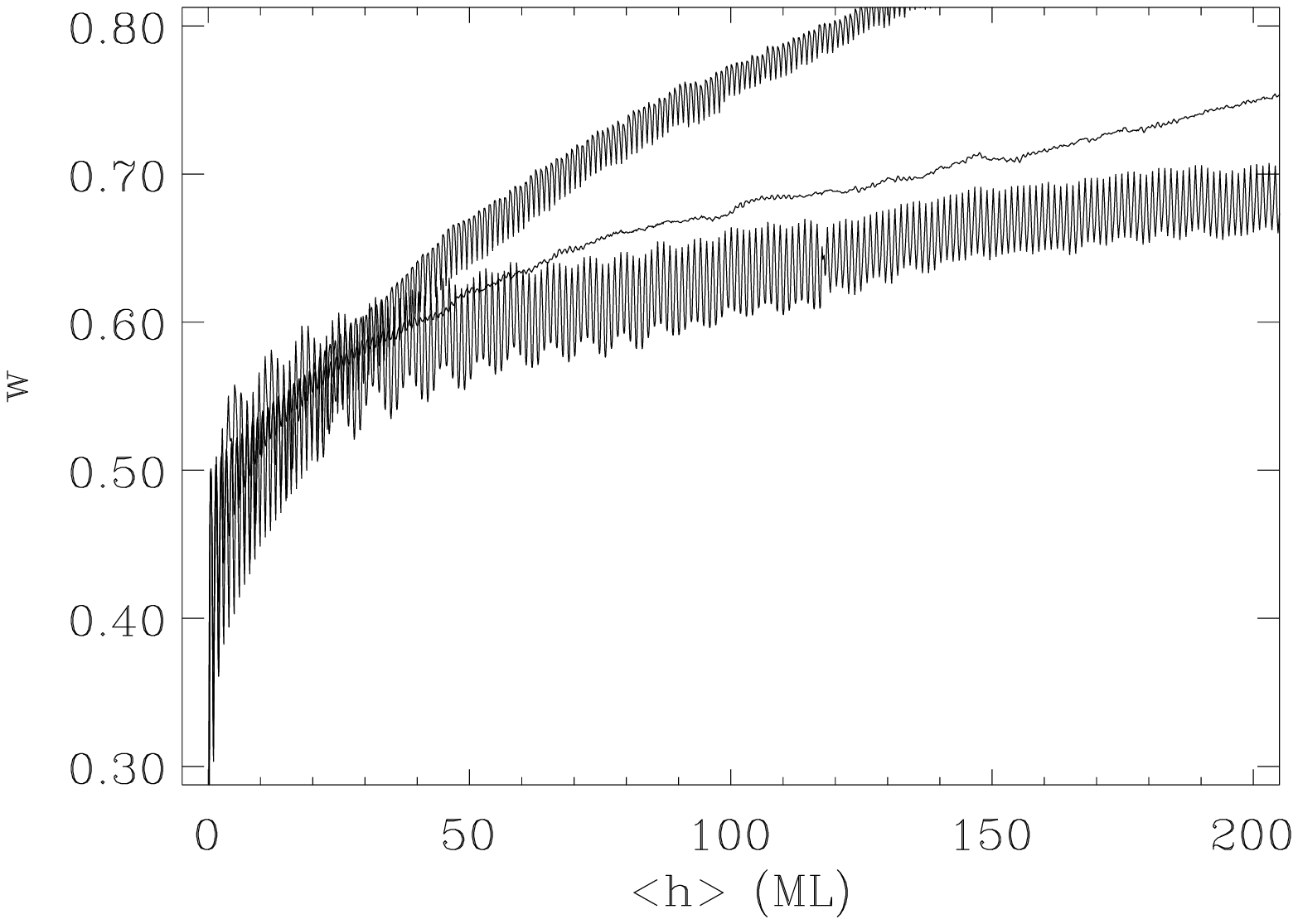,width=\figwidth}}}
\put(0.4,0.083){\parbox[b]{0.417\figwidth}{\psfig{file=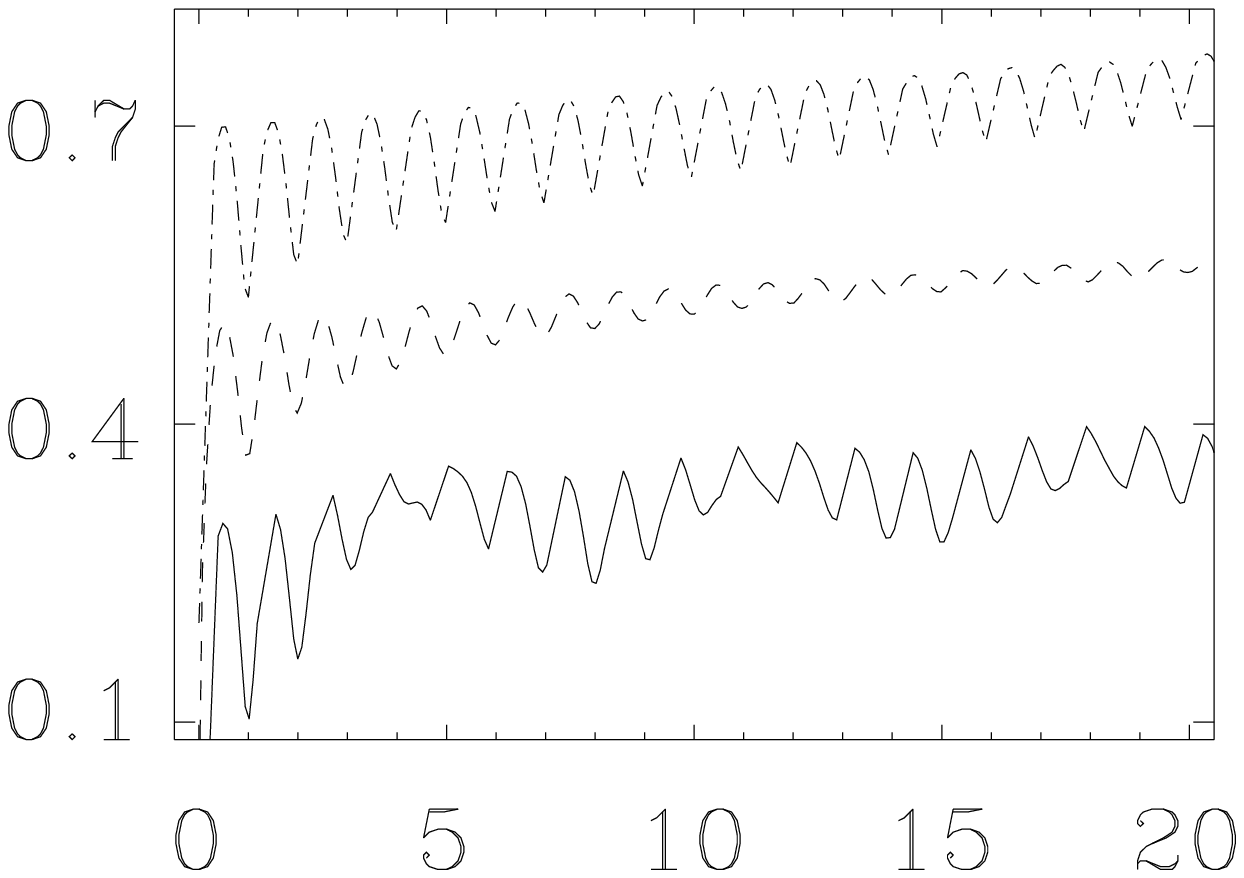,width=0.5\figwidth}}}
\put(0.8,0.275){\mbox{B}}
\put(0.68,0.31){\mbox{A}}
\put(0.57,0.36){\mbox{C}}

\put(0.88,0.545){\mbox{B}}
\put(0.65,0.56){\mbox{A}}
\put(0.44,0.6){\mbox{C}}
\end{picture}
\caption{ \label{flush-normal} 
  Comparison of the surface width for conventional growth (A) and the
  flush--technique as described in the text. We consider the
  flush--technique with (B) and without desorption (C). The inset
  shows the surface width oscillations during the deposition of the
  first 20 ML.  For clarity of presentation we have shifted the upper
  (lower) curve about +0.2 (-0.2) inside the inset.  These simulations
  were carried out on a $512 \times 512$ lattice at 560 K. The results
  are averaged over six independent simulations runs in each case.
  }
\end{figure}

In addition to higher growth rates, layer--by--layer growth is
assisted by the flush--mode. In fig.\ \ref{flush-normal} we compare
three different techniques/models of growth:
\begin{itemize}
\item[A)] conventional growth with $F=1$ ML/s and allowed desorption,
\item[B)] flush--mode with $F_{\mbox{\small const}} = 1$ ML/s and
  additional 0.30 ML in 0.003 s each second and allowed desorption,
\item[C)] flush--mode without desorption ($E_D=\infty$),
  $F_{\mbox{\small const}} = 0.77$ ML/s and additional 0.23 ML during
  0.003 s each second.
\end{itemize}
The different fluxes in B and C are chosen in order to achieve the
synchronization of the pulses with layer--completion. Due to the
possible desorption in B, however, synchronization can be achieved
only approximately in this case.

We investigate these different methods by comparing the {\em surface
  width}
\begin{equation}
   w = \sqrt{\langle (h(x,y)-\langle h \rangle)^2 \rangle}.
\end{equation}
Perfect layer--by--layer growth would lead to oscillations between
zero and 0.5 (coverage of half a monolayer). Higher values of $w$
indicate a broader distribution of the heights.

After the deposition of 60 ML using the different techniques the
surface widths
become considerably different (see fig.\ \ref{flush-normal}).   The
flush--mode without desorption (C) is even farer away from perfect
layer--by--layer growth compared to conventional growth (A). The
flush--mode in the presence of desorption (B) is superior to both, (A)
and (C) in the long run and  keeps
the surface smooth. Looking at the deposition of the first 20 ML this
seems to be surprising. The oscillations of $w$ with technique (B) are
disrupted due to an obvious asynchronization. With (C) the
synchronization is perfect leading to very strong and regular
oscillations. Using the conventional technique (A) the
oscillations are damped and much less pronounced.

To summarize, the usage of the proposed flush--technique is useful to
improve the growth rate (we achieved a factor two at high $T$) and to
assist layer--by--layer growth. Hereby, the desorption of adatoms is
crucial to achieve optimized growth: without desorption the
flush--mode is worse compared to conventional growth even though
strong oscillations are induced. The reevaporation has such an impact
because it is {\em height selective}, {\it i.e}.\ adatoms on top of
existing islands desorb easily whereas adatoms beneath islands
preferentially are incorporated in the crystal.  Clearly, this height
selective behaviour is achieved only when a positive
Ehrlich--Schwoebel barrier hinders the particles to be incorporated at
step edges from above.

We would like to point out that the usage of a chopped flux has been
proposed and investigated by Rosenfeld {\em et al}.\ \cite{rlw95} in
the framework of the {\em concept of the two mobilities}. However, our
findings show that only in the presence of desorption the occurrence
of oscillations is indeed coupled to a reduction of the surface
roughness. In ref.\ \cite{jn97} the effect of a chopped flux on island
distances was investigated. These findings would allow to optimize
even further in that one calculates the minimal flux intensity and the
time of the flush needed in order to achieve an increased island
density. Here, we have chosen a safe high flux without explicit use of
the results of \cite{jn97}.

\section{Optimizing the structure of mounds in 3D--growth}  \label{sed}
Quite generally, layer--by--layer growth \cite{kbk97}, as well as step
flow is not attainable forever \cite{ks95,rsk96}.  This can be due to
{\em e.g.} Ehrlich--Schwoebel barriers \cite{vil91} which is
typically positive \cite{sto94}. This favors new
nucleation events on top of existing islands which leads to 3D--growth
sooner or later. In order to optimize MBE growth it is thus also
interesting to study the growth of 3D--structures by computer
simulations.

We will start with a brief summary of our findings for 3D--growth on
the basis of a simplified model of epitaxial growth
\cite{bks98,skb98}. This will enable us to introduce the basic
concepts. After that, we will show how these results can be used in
order to improve 3D--growth (to be specified below). We will test this
new strategy with computer simulations of the SOS--model of Sec.\
\ref{model}.

The most important simplification we introduced in ref.\ \cite{bks98}
was an effective description of diffusion and nucleation. Rather than
to simulate the simultaneous motion of many adatoms we concentrated on
the simulation of individual particles which is a usual technique for
simple growth models \cite{wv90,dt91}. Parameters to the model are the
diffusion length and in a similar way SED is considered. Even though a
similar SED has been introduced in ref.\ \cite{be95}, there, in
difference to the present work no search for kink sites was
implemented. 

In MBE the typical length of the step edge diffusion process depends
on temperature and flux of arriving particles \cite{vp95}. On a
one--dimensional substrate the theory of island nucleation predicts a
typical distance between nucleation centers of the form
\begin{equation}  \label{flsed}
  \lsed \approx \left( \frac{d}{f} \right)^{1/4}
\end{equation}
where $d$ is the diffusion constant and $f$ the flux of arriving
particles \cite{wol97}.  If we apply this theory to the lateral or
in--plane growth of a pyramid (concentrating on a slice of one ML
thickness) the flux $f$ can be identified with the reduced flux per
unit length of the step edge $f=F \ell_T$ where $\ell_T$ stands for
the terrace width. Within this context $d$ becomes the diffusion
constant for diffusion along the step edge. The scaling relation
(\ref{flsed}) for \lsed\ was obtained under the restriction that two
atoms (i.e. $i^* + 1=2$) already form a stable nucleus and no
desorption occurs. We note that for greater values of $i^*$ the
correct theoretical result has been derived recently
\cite{kw98}. However, for the model as described in Sec.~\ref{model}
the assumption of $i^*=1$ is reasonable.

When \lsed\ is of the order of the modeled system size (strong SED)
the growth is characterized by the formation of square based pyramids
with a well defined slope. The step edges are oriented along the
lattice coordinates and the surface width was found in Ref.\
\cite{bks98} to grow with a power law
\begin{equation}
   w \propto h^\beta \mbox{ with } \beta \approx 0.45
\end{equation}
where $\beta$ is called the growth exponent \cite{bar95}.  Typically,
one expresses the scaling behaviour in terms of the elapsed
time. However, in the context of this paper it is advantageous to use
the mean height $h$ instead (as will become clear soon).  The typical
distance between the pyramids (the correlation length) was found to be
proportional to
\begin{equation} \label{fxi}
 \xi \propto h^{1/z} \mbox{ and } z=\alpha/\beta \approx 2.3
\end{equation} 
in accordance to the occurrence of slope selection. More formally this
means that the ratio of the typical length scales, $w$ and $\xi$,
remains constant, and hence $\alpha=1$.

The relatively high growth exponent of 0.45 reflects the fact that the
coarsening process is SED--assisted \cite{skb98}. Due to the strong
SED material is moved efficiently towards regions with high densities
of kink sites, {\it i.e}.\ towards the contact points of pyramids or
mounds. For lower values of \lsed\ the coarsening process is purely
noise assisted \cite{tsv98}. Hence, the structures are merging more
slowly.

If the size of the pyramids exceeds \lsed\ the pyramids loose their
perfect shape. The structures become round, and step edges will be
fringed \cite{skb98}. It is clear that due to the coarsening,
conventional MBE growth is bound to drive itself into this state. 

\begin{figure}[t]
\setlength{\unitlength}{\figwidth}
\begin{picture}(1,0.707)(0,0)
\put(0,0){\parbox[b]{\figwidth}{\psfig{file=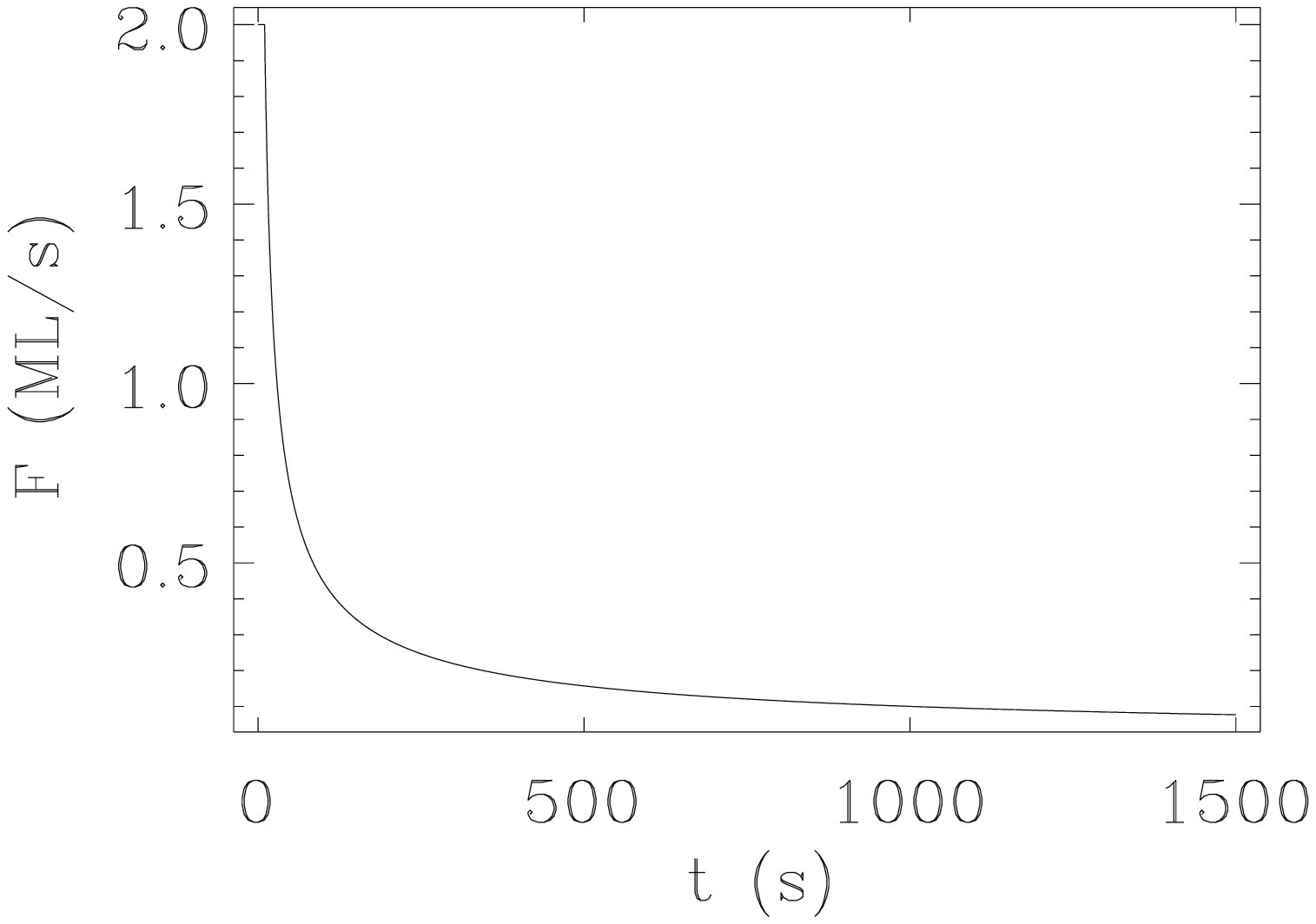,width=\figwidth}}}
\put(0.4,0.2){\parbox[b]{0.5\figwidth}{\psfig{file=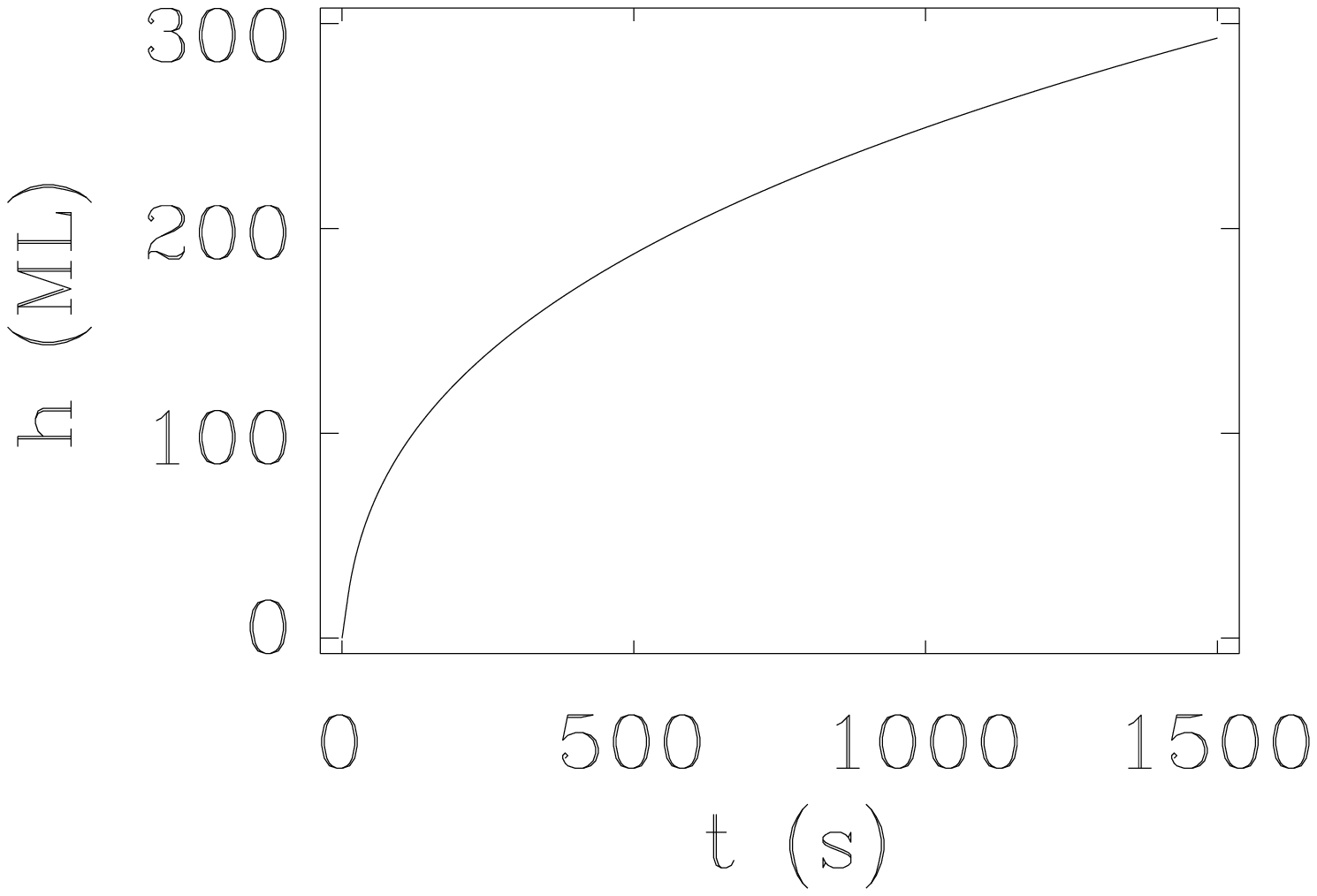,width=0.5\figwidth}}}
\end{picture}
\caption{ \label{skizze2}
  Flux variation used in the simulation of flux--adaption of
  fig.\ \ref{stat}. The inset shows the reached height in dependence
  of the time.
}
\end{figure}

Now, we turn to the investigation how the latter stage in MBE growth
can be prevented. The main idea is that in order to prevent the
occurrence of rough step edges one has to require that always $\lsed
\approx \xi$. In the following we demonstrate how to fulfill this
condition by varying the flux $F$ of arriving particles. Equally well,
one could adapt the growth temperature. However, for this a detailed
knowledge of the activation energy of $d$ and the temperature
dependence of $\ell_T$ \cite{ksb98} is necessary which is often not
available.

Equating the expressions (\ref{fxi}) and (\ref{flsed}) we obtain the
height dependence of the flux
\begin{equation} \label{ffvh}
F(h) = c  h^{-4/z}
\end{equation}
where $c$ is an adequate constant.  To reformulate this relation in
terms of the time we use $\mbox{d}h/\mbox{d}t = F$ and solve the resulting
differential equation, obtaining
\begin{equation} \label{fhvt}
h(t) \propto t^{-z/(4+z)}.
\end{equation}
Reinserting this result into (\ref{ffvh}) we obtain that the flux
should be varied according to
\begin{equation}  \label{ft}
F(t) \propto t^{-4/(4 + z)} \approx t^{-0.65}
\end{equation}
where we inserted $z=2.3$ according to SED--assisted coarsening
\cite{bks98}.

We applied this strategy to the growth of the SOS--model of Sec.\
\ref{model} at 560 K. Clearly, SED is not a process which is
explicitly considered in this model. Typically, the atoms with only a
single bond to a step edge will detach and diffuse on the
terrace. However, the net result will be the same: single bonded atoms
will be moved to places with higher coordination (kink sites).  To
prevent the inference of reevaporation we suppressed this process
\cite{srk98}. However, we checked that even with desorption, the
strategy is still applicable and useful. The flux was chosen as shown
in fig.\ \ref{skizze2}. We started with a constant flux of 2
ML/s. After the growth of twenty monolayers we adapted the flux with
time $t$ according to $F = F_0 / (t/10 \mbox{s})^{0.65}$.

\begin{figure}[t]
\psfig{file=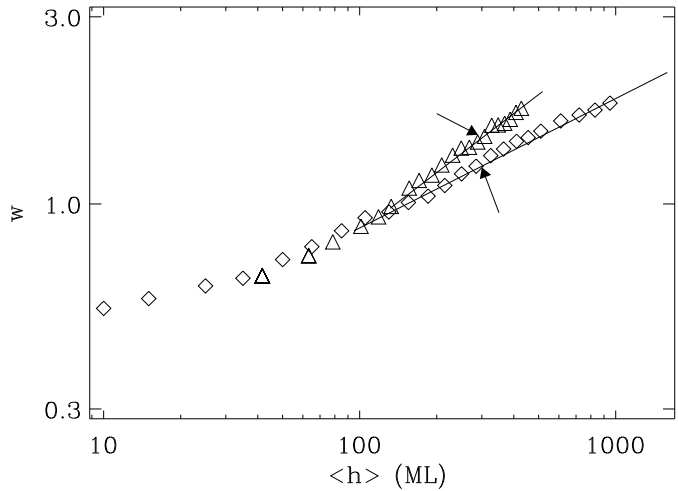,width=\figwidth}
\caption{                      \label{stat}
Comparison of the surface width under optimized growth conditions
$(\triangle)$ and without an adaption of the flux $F= 1$ ML/s
$(\Diamond)$.  Under optimized growth conditions the flux was
initially set to 2 ML/s and after 10 s adapted according $F = 2 /(t/10
\mbox{s})^{0.65}$ML/s (see fig.\ \ref{skizze2}). The solid lines
correspond to growth exponents $\beta = 1/3$ and $\beta=1/2$. The
arrows mark the positions of the snapshots of fig.\ \ref{surfaces}}
\end{figure}

\begin{figure}[t]
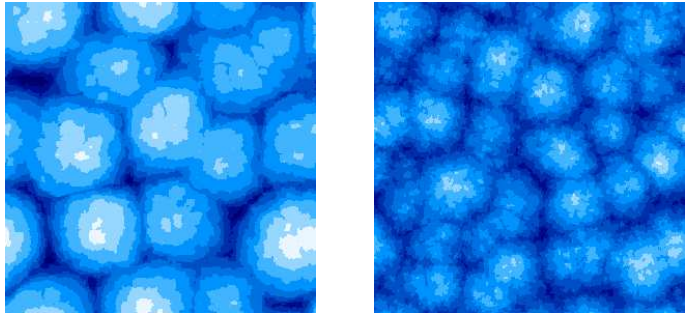

  \begin{center}
    \setlength{\unitlength}{\figwidth}
    \makebox(0.4,0.4){\psfig{file=surf07.ceps,width=0.4\figwidth}}
    \hspace{0.05\figwidth}
    \makebox(0.4,0.4){\psfig{file=surf37.ceps,width=0.4\figwidth}}
  \end{center}
\caption{\label{surfaces}
  Snapshots of surface morphology of $300 \times 300$ lattices where
  about 300 ML were deposited with the two different methods as in
  fig.\ \ref{stat} (left: flux--adaption, right: conventional growth).}
\end{figure}

In fig.\ \ref{stat} we compare the resulting evolution of the surface
width $w$ with a simulation with a constant flux $F=1$ ML/s. With the
new strategy we obtain a higher growth exponent of $\beta \approx 1/2$
compared to conventional growth with $\beta \approx 1/3$. These
exponents fit well to $\beta = 0.45$ for strong SED \cite{bks98} and
$\beta = 0.33$ for intermediate values of \lsed\ \cite{skb98}. The
result that $w$ grows fast and is described by a high growth exponent
under these optimized growth conditions should not however be confused
with the notion of a fast roughening, self--affine surface. It just
means that the structures are merging fast and the mounds are getting
high and wide. This can be seen directly in fig.\
\ref{surfaces}. After the deposition of 300 ML under constant flux the
structures are small whereas under optimized growth conditions the
resulting structures are larger. Note that because of the higher
initial flux of 2 ML/s the island density was much higher in the
beginning under optimized growth conditions. Nevertheless the
SED--assisted coarsening leads to a considerably fewer number of
mounds (approximately 10 which should be compared to 20 with the
conventional growth).

It is clear that in order to obtain larger structures in conventional
MBE one just has to grow for longer times. The step edges will become
smooth due to the equilibration after the growth has been
stopped. However, during growth the step edges do not remain smooth as
in our optimized growth mode. Hence, a larger probability for the
creation of vacancies or other crystal faults will be present. After
growth stops these faults can probably be only partially eliminated in
the non--optimized growth. We also mention that in the end another
processes will of course become important too. In the limit of $t
\rightarrow \infty$ no net growth will be achieved in the optimized
growth and the equilibration of the surface will be dominant.

\section{Conclusion}                  \label{conclusion}
We have investigated the effect of the microscopic dynamics on
experimentally accessible macroscopic effects in MBE growth. Based on
simulations of the solid--on--solid model we proposed two optimized
growth--strategies. 

Comparing the layer--by--layer growth with sublimation we understood
how the desorption of single adatoms comes into play during
growth. During growth, freely diffusing adatoms are created by the
external flux. During sublimation, however, such adatoms must first be
created, {\it e.g}.\ through detachment from steps. This difference
manifests itself in the different contribution of the microscopic
activation energies to the effective energies. The diffusion barrier
$E_B$ increases the effective energy of sublimation whereas it
decreases the activation energy of the desorption rate during growth.
Since the macroscopic desorption rate is influenced by the typical
lifetime of single atoms we are able to intervene in the growth
process. We showed that a flush--mode is able to prolong the
layer--by--layer growth regime and to reduce the desorption rate:
applying short pulses of particles we create a high density of
islands. Afterwards with a low flux one completes the monolayer. At
least for our particular simulations the desorption was crucial to
obtain improved growth. Even though the flush--mode always induces
strong oscillations, only in combination with desorption it leads to
an improved growth. This can be explained with the height--selective
behaviour of desorption, {\it i.e}.\ desorption occurs preferentially
on top of islands as long as a positive Ehrlich--Schwoebel barrier is
present.

In experiments one should be able to produce such short flushes using
a chopper or pulsed laser deposition in conjunction with conventional
MBE. The method should be very useful in order to grow planar coherent
thin films, {\it e.g}.\ for application in quantum well structures.
In addition, such experiments would allow to decide wether desorption
occurs out of a physisorbed precursor which has been debated in the
literature \cite{blw95,tnz97}. If so, one should not obtain an
improved growth rate using the flush--mode.

However, since layer--by--layer growth is unstable, the optimization
of 3D--growth might be useful too. Based on recent results concerning
the step edge diffusion we proposed to vary the flux of arriving
particles in order to maintain smooth edges already during
growth. Reducing the flux according to $F \propto t^{-0.65}$ we were
able to recover the high growth exponent of $\beta \approx 0.45$
measured on the simplified MBE model \cite{bks98}. The fast coarsening
process (since SED assisted) yields structures which soon become very
large compared to those of conventional MBE. Irrespective of the
desired size of the structures they can be produced under the same
(strong SED) conditions which is accomplished by variation of
$F$. Otherwise the MBE growth would drive itself in the regime where
\lsed\ is less then the typical extension of the structures. Thus, our
method opens new possibilities for the controlled creation of these
selforganized nanostructures by MBE. In addition, this strategy should
reduce the probability for the creation of vacancies since during the
conventional growth the rough edges would be overgrown later.
However, this is speculative and cannot be verified in the framework
of the solid--on--solid model.

Typically, rather low fluxes are used in order to improve the quality
of the grown structures. However, our result suggests that it is not
disadvantageous to apply higher fluxes in the beginning. In the end,
when the resulting structures are rather large it becomes important to
reduce the flux in order to adapt it to the {\em smoothening range} of
the step edge diffusion.

\begin{center} *** \end{center}
This work is supported by the Deutsche Forschungsgemeinschaft through
Sonderforschungsbereich 410.

\bibliography{Literatur}

\begin{thebibliography}{10}
\bibitem{note2}
  The sublimation energy can be derived from the oscillation period
  {\em e.g}.\ of the width of the height distribution, for short: the
  surface width $w$.
\end{thebibliography}
\bibliographystyle{unsrt}

\end{document}